\documentstyle[seceq,epsf,preprint]{ptptex}
\newcommand{\bra}[1]{\langle {#1} |}     %%
\newcommand{\ket}[1]{| {#1} \rangle}     %%
\newcommand{\be}{\begin{equation}}
\newcommand{\ee}{\end{equation}}
\newcommand{\ba}{\begin{array}}
\newcommand{\ea}{\end{array}}
\newcommand{\bc}{\begin{center}}
\newcommand{\ec}{\end{center}}

\newcommand{\prb}[1]{Phys. Rev. {\bf #1}}

\newcommand{\npb}[1]{Nucl. Phys. {\bf #1}}
\newcommand{\la}{\langle}
\newcommand{\ra}{\rangle}

\title{%        %You can use \\ for explicit line-break
Time-Evolution of a Collective Meson Field by the use of 
a Squeezed State
}
\author{%       %Use \sc for the family name
Yasuhiko {\sc Tsue}, Akinori {\sc Koike} and Naoko {\sc Ikezi}$^*$ 
}

\inst{%         %Affiliation, neglected when [addenda] or [errata]
Physics Division, Faculty of Science, Kochi University, Kochi 780-8520, 
Japan\\
$^{*}$Department of Physics, Nagoya University, Nagoya 464-8602, Japan
}

\recdate{%      %Editorial Office will fill in this.
}

\abst{%       %this abstract is neglected when [addenda] or [errata]
A time-evolution of quantum meson fields is investigated in 
a linear sigma model by means of the time-dependent variational 
approach with a squeezed state. 
The chiral condensate, which is a mean field of the quantum 
meson fields, 
and quantum fluctuations around it 
are treated self-consistently in this approach. The attention 
is payed to the description of the relaxation process of 
the chiral condensate, where the energy 
stored in the mean field configuration flows to the fluctuation 
modes. It is shown that the quantum fluctuations plays an important 
role to describe this relaxation process. 
}
\begin{document}

\maketitle

\section{Introduction}

One of the recent interests associated with the physics of 
the relativistic heavy-ion collisions is the investigation of the 
dynamics of matter at very high energy density. 
As for the chiral phase transition, especially, 
the possible formation of a disoriented chiral 
condensate in ultra-relativistic nuclear collisions has been 
discussed.\cite{BLAIZOT}\tocite{ISHIHARA} 
The time-evolution of the chiral condensate in the 
chiral phase transition is a main subject to study in diverse work, 
namely, the dynamics of the phase transition is investigated. 
In these studies, the O(4) sigma model has been used mainly as an 
effective model of quantum chromodynamics (QCD). 
The time-evolution of a scalar field is also important 
to describe some aspects of inflationary model of 
universe.\cite{GUTH}\tocite{MOTTOLA} 
However, most analytical results in this field concern 
the classical equations of motion. 

When the high energy nucleus-nucleus collisions occur, one 
expects the formation of the quark-gluon plasma. Also, 
it is expected that the chiral phase transition occurs and 
the chiral symmetric phase is realized. After the collisions, 
the hot region created by the collision will expand spatially and 
the temperature will become lower. Thus, 
the chiral phase transition will occurs again and 
the chiral broken phase will be realized. 
Here, let us consider so-called quench scenario. 
The chiral condensate which is a chiral order parameter 
must grow from 0 to $\la\sigma\ra$. The relaxation process 
should be realized since the energy stored initially 
should flow elsewhere. 
Thus, the interesting and important problem is to make it 
clear how the relaxation occurs. 
The dissipative process in the linear sigma model 
has been treated in the framework 
of the Caldeira-Leggett theory\cite{YABU} in the context 
of the decay of the disoriented chiral condensate. 
In their work, the mechanism of relaxation 
has been given within the classical level.

One of the present authors (Y.T.) together with 
D. Vautherin and T. Matsui has indicated that the quantum 
fluctuation plays an important role for the chiral phase 
transition\cite{TVM1} and the relaxation of the 
collective isospin rotation of the chiral condensate.\cite{TVM2} 
The Gaussian approximation of the functional Schr\"odinger 
picture\cite{JK} for the field theory has been adopted 
in our previous work. 
In this approach, both the mean field and the quantum fluctuations 
can be treated self-consistently including the quantum effect 
of the higher order of $\hbar$.
This approach is essentially same as 
the time-dependent variational approach with a squeezed 
state.\cite{TF} 
In this paper, we will use the later approach to investigate the 
time-evolution of the chiral condensate and the quantum fluctuation 
around it because the fluctuation modes can be constructed 
explicitly. 
Except for the previous work,\cite{TVM1,TVM2} 
the time-development of the chiral condensate 
was examined to leading order in a large $N$ expansion 
in Ref.\citen{COOPER} by Cooper et. al. 
However, in our time-dependent variational approach 
with a squeezed state, it is easily possible to investigate 
the role of the quantum fluctuations systematically 
because the quantum state is definitely constructed by means of
the squeezed state and the 
mode-expansion for quantum fluctuations is 
explicitly performed. 
The main aim of this paper is to point out that the quantum 
fluctuations are responsible for the occurrence of 
the relaxation process of the chiral condensate or 
chiral order parameter. Of course, the expansion of the system 
leads to a friction term.\cite{RUNDRAP,ISHIHARA} 
However, even if the expansion is not taken 
into account, it will be shown that 
the behavior like the damped oscillation for 
the chiral condensate appears due to the inclusion of the 
quantum fluctuations. 

This paper is organized as follows : 
The time-dependent variational approach with a squeezed state 
is reviewed briefly and the mode functions in this approach are 
introduced in the case of the uniform condensate in the next section. 
This approach is applied to the linear sigma model and the basic equations 
are obtained in \S 3. 
The expanding geometry is presented in Appendix. 
The numerical results and discussion are given in \S 4. The last 
section is devoted to a summary.

\section{Time-dependent variational approach with a squeezed state}

In this section we briefly give the basic ingredients in 
the time-dependent variational approach with a squeezed 
state.\cite{TF} 
The main advantage of this approach to the dynamical problem 
including both the degrees of freedom of the mean field (order 
parameter) and the quantum fluctuations around it 
(``mesons" if quantized) 
is in the fact that both degrees of freedom can be treated 
self-consistently through the nonlinear coupling. Further, 
we can get the canonical equations of motion describing 
their time-development.

\subsection{Basic equations of motion}
Apart from the sigma model, for simplicity, 
let us first consider a scalar field theory with one component. 
We start with the following Hamiltonian : 
%
%%%%%%%%%%%%%%%%%%%%%%%%%%%%% (2.1) %%%%%%%%%%%%%%%%%%%%%%%%%%%%%%%
\begin{equation}\label{2-1}
  {\hat H} = 
     \int d^3{\mib x} 
    \left\{ \frac{1}{2} \pi({\mib x})^2+ \frac{1}{2}
     \left(\nabla \phi({\mib x})\right)^2 + U[\phi({\mib x})] 
   \right\} 
      \ . 
\end{equation}
%%%%%%%%%%%%%%%%%%%%%%%%%%%%%%%%%%%%%%%%%%%%%%%%%%%%%%%%%%%%%%%%%%
%
We take a following squeezed state as a trial state :
%
%%%%%%%%%%%%%%%%%%%%%%%%%%%%% (2.2) %%%%%%%%%%%%%%%%%%%%%%%%%%%%%%%
\begin{eqnarray}\label{2-2}
  \ket{\Phi(t)} 
   &=&
    \exp \left(S(t)\right) \cdot 
    N(t)\exp \left(T(t)\right) \ket{0} \ , \\
 S(t) 
     &=& i\int d^3{\mib x} 
      [{\overline \pi}({\mib x},t)\phi({\mib x}) 
     - {\overline \varphi}({\mib x}, t) \pi({\mib x}) 
         ] \ , \nonumber\\
 T(t) 
     &=& \int\int d^3{\mib x} d^3{\mib y}
      \ \phi({\mib x}) \left[-\frac{1}{4}
      (G^{-1}({\mib x},{\mib y},t)-G^{(0)}({\mib x},{\mib y})^{-1})
      +i\Sigma({\mib x},{\mib y},t)\right]
      \phi({\mib y}) \ , \nonumber
\end{eqnarray}
%%%%%%%%%%%%%%%%%%%%%%%%%%%%%%%%%%%%%%%%%%%%%%%%%%%%%%%%%%%%%%%%%%
%
where $\ket{0}$ is the reference vacuum with respect to free boson 
annihilation operators $a_k$; $a_k \ket{0}=0$, and 
$G^{(0)}({\mib x},{\mib y})=\bra{0} \phi({\mib x})\phi({\mib y}) \ket{0}$. 
%Here the 
%Schr\"odinger representation is adopted. 
Further, $N$ is a 
normalization factor. The variational functions are 
${\overline \varphi}({\mib x},t)$, ${\overline \pi}({\mib x},t)$, 
$G({\mib x},{\mib y},t)$ and $\Sigma({\mib x},{\mib y},t)$, 
respectively.
The expectation values for the field operators are obtained as 
%
%%%%%%%%%%%%%%%%%%%%%%%%%%%%% (2-3) %%%%%%%%%%%%%%%%%%%%%%%%%%%%%%%
\begin{eqnarray}\label{2-3}
  &\ & \bra{\Phi(t)}\phi({\mib x})\ket{\Phi(t)} 
       = {\overline \varphi}({\mib x},t) \ , \qquad
  \bra{\Phi(t)}\pi({\mib x})\ket{\Phi(t)} 
       = {\overline \pi}({\mib x},t) \ , \nonumber\\
  &\ & \bra{\Phi(t)}\phi({\mib x})\phi({\mib y})\ket{\Phi(t)} 
       = {\overline \varphi}({\mib x},t){\overline \varphi}({\mib y},t) 
         + G({\mib x},{\mib y},t) 
       \ , \nonumber\\
  &\ & \bra{\Phi(t)}\pi({\mib x})\pi({\mib y})\ket{\Phi(t)} 
       = {\overline \pi}({\mib x},t){\overline \pi}({\mib y},t) 
         + \frac{1}{4}G^{-1}({\mib x},{\mib y},t)
        +4\bra{{\mib x}}\Sigma(t)G(t)\Sigma(t)\ket{{\mib y}} \ ,\nonumber\\
\end{eqnarray}
%%%%%%%%%%%%%%%%%%%%%%%%%%%%%%%%%%%%%%%%%%%%%%%%%%%%%%%%%%%%%%%%%%
%
where we used the following notation :
%
%%%%%%%%%%%%%%%%%%%%%%%%%%%%% (2-4) %%%%%%%%%%%%%%%%%%%%%%%%%%%%%%%
\begin{equation}\label{2-4}
  \bra{\mib x}\Sigma(t)G(t)\Sigma(t)\ket{\mib y} \equiv
  \int\int d^3{\mib x}' d^3{\mib y}' 
  \Sigma({\mib x},{\mib x}',t)G({\mib x}',{\mib y}',t)
  \Sigma({\mib y}',{\mib y},t) \ .
\end{equation}
%%%%%%%%%%%%%%%%%%%%%%%%%%%%%%%%%%%%%%%%%%%%%%%%%%%%%%%%%%%%%%%%%%
%
Here, ${\overline \varphi}({\mib x},t)$ represents 
the condensate or the mean field. 
The two-point function $G({\mib x},{\mib y},t)$ means the two-point 
correlation function with the order of $\hbar$. This function 
represents the quantum fluctuation around the mean field 
${\overline \varphi}({\mib x},t)$. 
For later convenience, we introduce 
${\cal M}^{(n)}[{\overline \varphi}({\mib x},t)]$ defined as 
%
%%%%%%%%%%%%%%%%%%%%%%%%%% (2-5) %%%%%%%%%%%%%%%%%%%%%%%%%%%%%%
\begin{equation}\label{2-5}
{\cal M}^{(n)}[{\overline \varphi}({\mib x},t)]
   =\exp\left\{\frac{1}{2}
           G({\mib x},{\mib x},t)
              \frac{\partial^2}{\partial z^2}
     \right\}
     U^{(n)}[z] \biggl|_{z={\overline \varphi}({\bf x},t)} \ 
\end{equation}
%%%%%%%%%%%%%%%%%%%%%%%%%%%%%%%%%%%%%%%%%%%%%%%%%%%%%%%%%%%%
%
with $U^{(n)}[z]={d^n U}/{dz^n}$.
Here, ${\cal M}^{(0)}[{\overline \varphi}({\mib x},t)]$ 
is nothing but 
the expectation value of the functional $U[\phi({\mib x})]$
of the field operator $\phi({\mib x})$. 
From this expression, it is seen that the fluctuation around the mean
field can be taken into account with the loop contributions 
$G({\mib x}, {\mib x}, t)$.

The state $\ket{\Phi(t)}$ is determined as an functional of 
${\overline \varphi}({\mib x},t)$,
${\overline \pi}({\mib x},t)$, $G({\mib x},{\mib y},t)$ and 
$\Sigma({\mib x},{\mib y},t)$,
and their time-development is governed by the time-dependent variational 
principle :
%
%%%%%%%%%%%%%%%%%%%%%%%%%%%%% (2-6) %%%%%%%%%%%%%%%%%%%%%%%%%%%%%%%
\begin{equation}\label{2-6}
  \delta \int dt \bra{\Phi(t)} i\frac{\partial}{\partial t} - {\hat H}
                 \ket{\Phi(t)} = 0 \ .
\end{equation}
%%%%%%%%%%%%%%%%%%%%%%%%%%%%%%%%%%%%%%%%%%%%%%%%%%%%%%%%%%%%%%%%%%
%
The expectation value of the Hamiltonian is expressed as 
%
%%%%%%%%%%%%%%%%%%%%%%%%%%%%% (2-7) %%%%%%%%%%%%%%%%%%%%%%%%%%%%%%%
\begin{eqnarray}\label{2-7}
  & &H= \bra{\Phi(t)} {\hat H} \ket{\Phi(t)} 
   =  \int d^3{\mib x} {\cal E}({\mib x}) \ , \\
 & &{\cal E}({\mib x}) 
   = \frac{1}{2}{\overline \pi}({\mib x},t)^2
      +\frac{1}{2}\left(
       \nabla{\overline \varphi}({\mib x},t)\right)^2
      +{\cal M}^{(0)}[{\overline \varphi}({\mib x},t)] \nonumber\\
   & &\qquad \quad +\frac{1}{8}\bra{\mib x}G^{-1}(t)\ket{\mib x}
        +2\bra{\mib x}\Sigma(t)G(t)\Sigma(t)\ket{\mib x}
        +\frac{1}{2} \lim_{{\bf y}\rightarrow {\bf x}} 
         \nabla_{\bf x}\nabla_{\bf y} \bra{\mib x}G(t)\ket{\mib y} 
     \ . \nonumber     
\end{eqnarray}
%%%%%%%%%%%%%%%%%%%%%%%%%%%%%%%%%%%%%%%%%%%%%%%%%%%%%%%%%%%%%%%%%%
%
The variational equation (\ref{2-6}) gives the canonical 
equations of motion :
%
%%%%%%%%%%%%%%%%%%%%%%%%%%%%% (2-8) %%%%%%%%%%%%%%%%%%%%%%%%%%%%%%%
\begin{eqnarray}\label{2-8}
& &  \frac{\partial {\overline \varphi}({\mib x},t)}{\partial t}
    =\frac{\delta H}{\delta {\overline \pi}({\mib x},t)} \ , \qquad\qquad
    \frac{\partial {\overline \pi}({\mib x},t)}{\partial t}
    = -\frac{\delta H}{\delta {\overline \varphi}({\mib x},t)} \ , 
      \nonumber\\
& &  \frac{\partial G({\mib x},{\mib y},t)}{\partial t}
  = \frac{\delta H}{\delta \Sigma({\mib x},{\mib y},t)} \ , \qquad
     \frac{\partial \Sigma({\mib x},{\mib y},t)}{\partial t}
    = -\frac{\delta H}{\delta G({\mib x},{\mib y},t)} \ .
\end{eqnarray}
%%%%%%%%%%%%%%%%%%%%%%%%%%%%%%%%%%%%%%%%%%%%%%%%%%%%%%%%%%%%%%%%%%
%
As a result, we obtain the following equations of motion :
%
%%%%%%%%%%%%%%%%%%%%%%%%%%%%% (2.9) %%%%%%%%%%%%%%%%%%%%%%%%%%%%%%%
\begin{eqnarray}
   & & \frac{\partial^2 {\overline \varphi}({\mib x},t)}{\partial t^2}
    -\nabla^2 {\overline \varphi}({\mib x},t) 
    +{\cal M}^{(1)}[{\overline \varphi}({\mib x},t)]=0
     \ , 
    \label{2-9}\\
   & & \left\{-\nabla_{\mib x}^2
     +{\cal M}^{(2)}[{\overline \varphi}({\mib x},t)]\right\}
       G({\mib x},{\mib y},t) 
      =\frac{1}{4}\bra{\mib x}G^{-1}(t)\ket{\mib y}
       -2\bra{\mib x}{\dot \Sigma}(t)G(t)\ket{\mib y}
       \nonumber\\
   & &\qquad\qquad\qquad\qquad\qquad\qquad\qquad\qquad
      -4\bra{\mib x}\Sigma(t)^2G(t)\ket{\mib y}
     \ , 
    \label{2-10}\\
   & & \frac{\partial G({\mib x},{\mib y},t)}{\partial t}
     =2[\bra{\mib x}G(t)\Sigma(t)\ket{\mib y}
                      +\bra{\mib x}\Sigma(t)G(t)\ket{\mib y}]
     \ ,
    \label{2-11}
\end{eqnarray}
%%%%%%%%%%%%%%%%%%%%%%%%%%%%%%%%%%%%%%%%%%%%%%%%%%%%%%%%%%%%%%%%%%
%
where we used the equation 
${\partial_t{\overline \varphi}}={\overline \pi}$. 
The above-derived equations of motion are the basic equations 
in the squeezed state approach.

\subsection{Mode expansion}
From now on, let us concentrate our attention on the 
equations of motion for 
quantum fluctuations in Eqs.(\ref{2-10}) and (\ref{2-11}).
We assume that the condensate is uniform spatially with a translational
invariance. With this assumption, the two-point functions 
$G({\mib x},{\mib y},t)$ and $\Sigma({\mib x},{\mib y},t)$ only depend 
on ${\mib x}-{\mib y}$. Thus it is possible to carry out the 
Fourier transformation as follows :
%
%%%%%%%%%%%%%%%%%%%%%%%%%%%%% (2-12) %%%%%%%%%%%%%%%%%%%%%%%%%%%%%%%
\begin{equation}\label{2-12}
    G({\mib x},{\mib y},t) 
  = \frac{1}{{(2\pi)^3}}\int d^3{\mib k} e^{i{\bf k}\cdot
       ({\bf x}-{\bf y})} G_{\bf k}(t) \ ,\quad 
    \Sigma({\mib x},{\mib y},t) 
  = \frac{1}{{(2\pi)^3}}\int d^3{\mib k} e^{i{\bf k}\cdot
        ({\bf x}-{\bf y})} 
        \Sigma_{\bf k}(t) \ . 
\end{equation}
%%%%%%%%%%%%%%%%%%%%%%%%%%%%%%%%%%%%%%%%%%%%%%%%%%%%%%%%%%%%%%%%%%
%
From Eq.(\ref{2-11}) the equation ${\dot G}_{\bf k} = 4 G_{\bf k} 
\Sigma_{\bf k}$ is obtained. 
By using the above transformation and Eq.(\ref{2-10}), 
we can get the following equation of motion for the Fourier modes 
$G_{\bf k}(t)$ :
%
%%%%%%%%%%%%%%%%%%%%%%%%%%%%%% (2-14) %%%%%%%%%%%%%%%%%%%%%%%%%%%%%
\begin{equation}\label{2-14}
    \frac{1}{2}{\ddot G}_{\bf k}(t) 
    - \frac{{\dot G}_{\bf k}(t)^2}{4G_{\bf k}(t)}
    + {\mib k}^2 G_{\bf k}(t) - \frac{1}{4 G_{\bf k}(t)} 
    + [{\cal M}^{(2)}G]_{\bf k} 
    = 0 \ ,
\end{equation}
%%%%%%%%%%%%%%%%%%%%%%%%%%%%%%%%%%%%%%%%%%%%%%%%%%%%%%%%%%%%%%%%%
%
where $[{\cal M}^{(2)}G]_{\bf k}$ means the Fourier-transformed 
mode of ${\cal M}^{(2)}[{\overline \varphi}(t)] 
G({\mib x},{\mib y},t)$ :\break 
${\cal M}^{(2)}[{\overline \varphi}(t)] G({\mib x},{\mib y},t) 
    = (1/(2\pi)^3)\int d^3{\mib k} 
    e^{i{\bf k}\cdot ({\bf x}-{\bf y})} [{\cal M}^{(2)}G]_{\bf k}$.
It should be noted here that $G_{\bf k}$ has the meaning of 
the Gaussian width in terms of the functional Schr\"odinger 
picture,\cite{JK,TF} that is, $G_{\bf k}$ is positive definite. 
We can thus express $G_{\bf k}(t)$ as a square of the mode function 
$\eta_{\bf k}$ :
%
%%%%%%%%%%%%%%%%%%%%%%%%%%%%% (2-15) %%%%%%%%%%%%%%%%%%%%%%%%%%%%%%%
\begin{equation}\label{2-15}
    G_{\bf k}(t) = \eta_{\bf k}(t)^2    \ . 
\end{equation}
%%%%%%%%%%%%%%%%%%%%%%%%%%%%%%%%%%%%%%%%%%%%%%%%%%%%%%%%%%%%%%%%%%
%
Substituting Eq.(\ref{2-15}) into Eq.(\ref{2-14}),
the equation of motion for fluctuation mode is obtained as 
%
%%%%%%%%%%%%%%%%%%%%%%%%%%%%% (2-16) %%%%%%%%%%%%%%%%%%%%%%%%%%%%%%%
\begin{equation}\label{2-16}
    {\ddot \eta}_{\bf k}(t) + {\mib k}^2 \eta_{\bf k}(t) 
    +[{\cal M}^{(2)}G]_{\bf k} \frac{1}{\eta_{\bf k}(t)}
    -\frac{1}{4 \eta_{\bf k}(t)^3} = 0 \ . 
\end{equation}
%%%%%%%%%%%%%%%%%%%%%%%%%%%%%%%%%%%%%%%%%%%%%%%%%%%%%%%%%%%%%%%%%%
%
The energy density in Eq.(\ref{2-7}) is expressed in terms of the mean field 
${\overline \varphi}(t)$ 
and the mode functions $\eta_{\bf k}(t)$ as follows :
%
%%%%%%%%%%%%%%%%%%%%%%%%%%%%% (2-17) %%%%%%%%%%%%%%%%%%%%%%%%%%%%%%%
\begin{equation}\label{2-17}
    {\cal E}({\mib x}) =
    \frac{1}{2}\left( 
     \frac{\partial {\overline \varphi}(t)}{\partial t} \right)^2
%    +\frac{1}{2}\left( 
%     \frac{\partial {\overline \varphi}(t)}{\partial x} \right)^2
    +{\cal M}^{(0)}[{\overline \varphi}(t)] 
%\nonumber\\
    +\frac{1}{{(2\pi)^3}}\int d^3{\mib k} \left\{ \frac{1}{2}
       {\dot \eta}_{\bf k}^2
        +\frac{1}{2}{\mib k}^2 \eta_{\bf k}^2 
        + \frac{1}{8 \eta_{\bf k}^2} \right\} \ . 
\end{equation}
%%%%%%%%%%%%%%%%%%%%%%%%%%%%%%%%%%%%%%%%%%%%%%%%%%%%%%%%%%%%%%%%%%
%

Here, we give two comments : One is about the mode expansion. 
If we rewrite the mode function into 
$\eta_{\bf k}(t)=1/\sqrt{2\epsilon_{\bf k}(t)}$, 
the two-point correlation function 
$G({\mib x},{\mib y},t)$ is expressed as 
$G({\mib x},{\mib y},t)=\sum_{\bf k}\xi_{\bf k}({\mib x},t)
\xi_{\bf k}^*({\mib y},t)/(2\epsilon_{\bf k}(t))$ where
$\xi_{\bf k}({\mib x},t)=e^{i{\bf k}\cdot {\bf x}}
e^{-i\int^t \epsilon_{\bf k}(t')dt'}$. 
This expression is nothing but that of the usual mode-expansion 
by the plane wave. Here $\epsilon_{\bf k}(t)$ corresponds to the 
time-dependent single particle energy. 
Another is the treatment of the system at finite temperature. 
If it is necessary to deal with the system at finite temperature 
in the thermal equilibrium state, 
the mode expansion must be slightly 
modified. In the thermal equilibrium, the variables are 
time-independent. 
The field operator is divided into 
$\phi={\overline \varphi}+{\hat \phi}$. 
Further, ${\hat \phi}$ is expanded by the mode functions :
\begin{equation}\label{2-18}
{\hat \phi}=\sum_{\bf k}\frac{1}{\sqrt{2\epsilon_{\bf k}}}
(\xi_{\bf k}b_{\bf k}+\xi_{\bf k}^* b_{\bf k}^{\dagger}) \ ,
\end{equation}
where $b_{\bf k}\ket{\Phi}=0$ at zero temperature. 
In the thermal equilibrium state, we assume that 
$\la b_{\bf k} b_{{\bf k}'}\ra=\la b_{\bf k}^{\dagger} 
b_{{\bf k}'}^{\dagger}\ra=0$ and $\la b_{\bf k}^{\dagger} 
b_{{\bf k}'}\ra=n_{\bf k}\delta_{{\bf k}{\bf k}'}$ where 
$n_{\bf k}$ is the bose distribution function : 
$n_{\bf k}=1/(e^{\beta\epsilon_{\bf k}}-1)$. 
Then, the two-point function 
$G({\mib x}, {\mib y})$ is expressed as 
\begin{equation}\label{2-19}
G({\mib x},{\mib y})=\sum_{\bf k}
\frac{1}{2\epsilon_{\bf k}}[\xi_{\bf k}({\mib x})\xi_{\bf k}^*({\mib y}) 
+(\xi_{\bf k}({\mib x})\xi_{\bf k}^*({\mib y}) +
\xi_{\bf k}^*({\mib x})\xi_{\bf k}({\mib y}))n_{\bf k}] \ . 
\end{equation}
Thus, it is necessary to replace the Fourier transformation in 
Eq.(\ref{2-15}) into 
%
%%%%%%%%%%%%%%%%%%%%%%%%%%%%% (2-20) %%%%%%%%%%%%%%%%%%%%%%%%%%%%%%%
\begin{equation}\label{2-20}
    G_{\bf k} = (1+2n_{\bf k})\eta_{\bf k}^2    \  
\end{equation}
%%%%%%%%%%%%%%%%%%%%%%%%%%%%%%%%%%%%%%%%%%%%%%%%%%%%%%%%%%%%%%%%%%
%
when we treat the system at finite temperature in the thermal 
equilibrium state.

\section{Squeezed state approach to the linear sigma model}

In this section, 
let us derive the equations of motion for the linear sigma 
model in this approach developed in \S 2. 
The Hamiltonian density is given as 
%
%%%%%%%%%%%%%%%%%%%%%%%%%%%%% (3-1) %%%%%%%%%%%%%%%%%%%%%%%%%%%%%%%
\begin{eqnarray}\label{3-1}
    {\cal H} 
    &=& 
      \frac{1}{2} \pi_{a}({\mib x})^2 
      +\frac{1}{2} \nabla \phi_{a}({\mib x}) \nabla\phi_{a}({\mib x})
      +\lambda\left(\phi_{a}({\mib x})^2-v^2 \right)^2 
         -h\phi_{0}({\mib x}) 
\end{eqnarray}
%%%%%%%%%%%%%%%%%%%%%%%%%%%%%%%%%%%%%%%%%%%%%%%%%%%%%%%%%%%%%%%%%%
%
where $a$ runs 0 $\sim N-1$. The index 0 
means the sigma-field 
and 1 $\sim N-1$ mean the pi-fields. 
Hereafter, we introduce $m^2= 4\lambda v^2$. 
% and 
%${\vec \phi}=(\phi_1, \phi_2, \phi_3)$. 
The trial state we here adopt is expressed as 
%
%%%%%%%%%%%%%%%%%%%%%%%%%%%%% (3-3) %%%%%%%%%%%%%%%%%%%%%%%%%%%%%%%
\begin{eqnarray}\label{3-3}
  & &\ket{\Phi(t)}
     =  \prod_{a=0}^{N-1}
     \exp\{S_{a}(t)\}\cdot N_{a}(t)\exp\{T_{a}(t)\}
     \ket{0} \ . 
\end{eqnarray}
%%%%%%%%%%%%%%%%%%%%%%%%%%%%%%%%%%%%%%%%%%%%%%%%%%%%%%%%%%%%%%%%%%
%
The expectation value of Hamiltonian density, 
$\bra{\Phi(t)}{\cal H}\ket{\Phi(t)}$, is easily calculated 
similar to Eq.(\ref{2-7}). 
%For example, 
%$\bra{\Phi(t)}{\cal H}_a \ket{\Phi(t)}$ has the same expression as
%${\cal E}({\mib x})$ in (\ref{2-7}) with the index $a$, where 
For later convenience, we define ${\cal M}_a^{(n)}$ same as 
Eq.(\ref{2-5}) : 
%
%%%%%%%%%%%%%%%%%%%%%%%%%%%%% (3-6) %%%%%%%%%%%%%%%%%%%%%%%%%%%%%%%
\begin{eqnarray}\label{3-6}
    {\cal M}_a^{(n)}[{\overline \varphi}_a({\mib x},t)] 
     &=& 
      \exp\left\{\frac{1}{2}G_a({\mib x},{\mib x},t)
       \frac{\partial^2}{\partial z^2}
          \right\}\cdot \frac{d^n}{dz^n}
          \left(-\frac{m^2}{2} z^2 +\lambda z^4\right)
          \biggl|_{z={\overline \varphi}_a({\bf x},t)} 
%          \nonumber\\ 
%     &=&
%       -\frac{m^2}{2}{\overline \varphi}_a({\mib x},t)^2 
%       +\lambda {\overline \varphi}_a({\mib x},t)^4 
%   +\frac{1}{2}G_{a}({\mib x},{\mib x},t)
%    \left(-m^2+12\lambda {\overline \varphi}_a({\mib x},t)^2\right)
%\nonumber\\
%    & &+3\lambda G_{a}({\mib x},{\mib x},t)^2 \ 
\end{eqnarray}
%%%%%%%%%%%%%%%%%%%%%%%%%%%%%%%%%%%%%%%%%%%%%%%%%%%%%%%%%%%%%%%%%%
%
with $a=0 \sim N-1$. 
%Also, 
%$\bra{\Phi(t)} \varphi_i({\mib x})^2\varphi_j({\mib x})^2
%\ket{\Phi(t)}= 
%({\overline \varphi}_{i}({\mib x},t)^2+G_{i}({\mib x},{\mib x},t)) 
%({\overline \varphi}_{j}({\mib x},t)^2+G_{j}({\mib x},{\mib x},t))
%$ 
%for $i\neq j$, which appears in the expectation values of 
%${\cal H}_{\sigma\pi}$ and ${\cal H}_{\pi\pi}$. 
We omit the constant term $m^4/(16\lambda)$ because of 
no contribution to the dynamics.

The time-dependent variational principle (\ref{2-6}) gives 
us the canonical equations of motion for 
$({\overline \varphi}_a , {\overline \pi}_a)$ and 
$(G_a , \Sigma_a )$ and leads us to the following equations :
%
%%%%%%%%%%%%%%%%%%%%%%%%%%%%% (3-7.8) %%%%%%%%%%%%%%%%%%%%%%%%%%%%%%%
\begin{eqnarray}
 \frac{\partial {\overline \varphi}_{a}({\mib x},t)}{\partial t}
    &=&\frac{{\delta H}}{\delta {\overline \pi}_{a}({\mib x},t)} 
    = {\overline \pi}_{a}({\mib x},t)\ , \nonumber\\
  \frac{\partial {\overline \pi}_{a}({\mib x},t)}{\partial t}
    &=&-\frac{{\delta H}}{\delta {\overline \varphi}_{a}({\mib x},t)} 
      \nonumber\\
  &=&
   -\biggl[-\nabla^2{\overline \varphi}_a({\mib x},t)
     +{\cal M}_a^{(1)}[{\overline \varphi}_a({\mib x},t)] \nonumber\\
  & &\ \ 
     +4\lambda{\overline \varphi}_a({\mib x},t)
     \sum_{b\neq a}\Bigl({\overline \varphi}_b({\mib x},t)^2
     +G_b({\mib x},{\mib x},t)\Bigl)-h\delta_{a0}\biggl]
     \ , \label{3-7}\\
   \frac{\partial G_{a}({\mib x},{\mib y},t)}{\partial t}
    &=& \frac{{\delta H}}{\delta \Sigma_{a}({\mib x},{\mib y},t)} 
    = 2\left[\bra{{\mib x}}G_{a}(t)\Sigma_{a}(t)\ket{{\mib y}}
          +\bra{{\mib x}}\Sigma_{a}(t)G_{a}(t)\ket{{\mib y}}\right]
     \ , \nonumber\\
  \frac{\partial \Sigma_{a}({\mib x},{\mib y},t)}{\partial t}
    &=& -\frac{{\delta H}}{\delta G_{a}({\mib x},{\mib y},t)} \nonumber\\
  &=&-\biggl[\frac{1}{2}
       {\cal M}_a^{(2)}[{\overline \varphi}_a({\mib x},t)]
       \delta({\mib x}-{\mib y})
       -\frac{1}{8}\bra{\mib x}\frac{1}{G_a(t)^2}\ket{\mib y}
       +2\bra{\mib x}\Sigma_a(t)^2\ket{\mib y} \nonumber\\
   & &\ \ 
%   \qquad\qquad\qquad\qquad
       +\frac{1}{2}\nabla_{\mib x}\nabla_{\mib y}\delta({\mib x}-{\mib y}) 
       +2\lambda\sum_{b\neq a}\Bigl( 
       {\overline \varphi}_b({\mib x},t)^2+G_b({\mib x},{\mib x},t)\Bigl)
       \delta({\mib x}-{\mib y})\biggl] \ .\nonumber\\
    \label{3-8}
\end{eqnarray}
%%%%%%%%%%%%%%%%%%%%%%%%%%%%%%%%%%%%%%%%%%%%%%%%%%%%%%%%%%%%%%%%%%
%
The last equation in Eq.(\ref{3-8}) is simply expressed 
by multiplying 
$G_a({\mib x},{\mib y},t)$ from the right-hand side : 
%
%%%%%%%%%%%%%%%%%%%%%%%%%%%% (3-9) %%%%%%%%%%%%%%%%%%%%%%%%%%%%%%
\begin{eqnarray}\label{3-9}
& &
\left\{-\nabla_{\mib x}^2 
+ {\cal M}_a^{(2)}[{\overline \varphi}_a({\mib x},t)] \right\}
G_a({\mib x},{\mib y},t)\nonumber\\
&=& \frac{1}{4}\bra{\mib x}G_a^{-1}(t)\ket{\mib y}
-2\bra{\mib x}{\dot \Sigma}_a(t)G_a(t)\ket{\mib y}
-4\bra{\mib x}\Sigma_a(t)^2G_a(t)\ket{\mib y} \nonumber\\
& & \ 
%\qquad\qquad\qquad\qquad\qquad\qquad\qquad\qquad
-4\lambda\sum_{b\neq a}\left({\overline \varphi}_b({\mib x},t)^2
+G_b({\mib x},{\mib y},t)\right)G_a({\mib x},{\mib y},t) \ .
\end{eqnarray}
%%%%%%%%%%%%%%%%%%%%%%%%%%%%%%%%%%%%%%%%%%%%%%%%%%%%%%%%%%%%%%%%
%

Let us consider the case where the system has translational 
invariance. 
Then, the condensates do not depend on space-coordinate 
${\mib x}$. 
Further, the two-point functions such as $G_a({\mib x},{\mib y},t)$ 
depend on 
the difference ${\mib x}-{\mib y}$ only. 
Then, we can carry out the Fourier transformation in the same 
way as those in the section 2. 
Comparing Eq.(\ref{2-10}) with Eq.(\ref{3-9}), 
we here regard ${\cal M}^{(2)}[{\overline \varphi}]$ in Eq.(\ref{2-10}) 
as ${\cal M}_a^{(2)}[{\overline \varphi}_a]+4\lambda\sum_{b\neq a}
[{\overline \varphi}_b^2+G_b(t)]$. 
Performing the mode expansion similar to Eqs.(\ref{2-12}) 
and (\ref{2-15}),  
we finally obtain the following equations of motion for the condensate 
(mean field) and the quantum fluctuations : 
%
%%%%%%%%%%%%%%%%%%%%%%%%%%%%% (3-10) %%%%%%%%%%%%%%%%%%%%%%%%%%%%%%%
\begin{eqnarray}\label{3-10}
  & &{\ddot {\overline \varphi}}_{a}(t)
     -m^2 {\overline \varphi}_{a}(t) 
     + 4\lambda {\overline \varphi}_{a}(t)^3
     +12\lambda \int\!\frac{d^3{\mib k}}{(2\pi)^3}
       \ \eta_{\bf k}^{a}(t)^2\cdot 
       {\overline \varphi}_{a}(t) \nonumber\\
  & & \qquad\qquad\qquad\qquad
     +4\lambda \sum_{b\neq a}\left({\overline \varphi}_{b}(t)^2 
     +\int\!\frac{d^3{\mib k}}{(2\pi)^3}\ \eta_{\bf k}^{b}(t)^2 \right)
     {\overline \varphi}_{a}(t) -h\delta_{a0} =0 \ , \nonumber\\
  & & {\ddot \eta}_{\bf k}^{a}(t) 
    + \biggl[{\mib k}^2-m^2+12\lambda {\overline \varphi}_{a}(t)^2 
    + 12\lambda\int\!\frac{d^3{\mib k}'}{{(2\pi)^3}}\  
    \eta_{{\bf k}'}^{a}(t)^2 
       \nonumber\\
  & & \qquad\qquad\qquad\qquad
     +4\lambda \sum_{b\neq a}\left({\overline \varphi}_{b}(t)^2 
     +\int\!\frac{d^3{\mib k}'}{(2\pi)^3}\ \eta_{{\bf k}'}^{b}(t)^2 
     \right) 
       \biggl]
    \eta_{\bf k}^{a}(t) 
    -\frac{1}{4 \eta_{\bf k}^{a}(t)^3} = 0 \ . \nonumber\\
\end{eqnarray}
%%%%%%%%%%%%%%%%%%%%%%%%%%%%%%%%%%%%%%%%%%%%%%%%%%%%%%%%%%%%%%%%%%
%
These are the basic equations in the linear sigma model in 
the squeezed state approach. The condensate and the quantum fluctuations 
are coupled each other and the time-evolution of both 
the degrees of freedom should be determined self-consistently.

It is instructive to deal with the static case in the 
O(4) linear sigma model. 
The differential terms with respect to time $t$ disappear 
in this case. 
Let us assume that ${\overline \varphi}_1={\overline \varphi}_2
={\overline \varphi}_3=0$, that is, the chiral condensate 
points in the sigma-direction. Also, it is assumed that 
the each fluctuation mode of the direction of pi is identical, which 
we denote $\eta^{\pi}$ : 
$\eta_{\bf k}^{1}=\eta_{\bf k}^{2}=\eta_{\bf k}^{3}=\eta_{\bf k}^{\pi}$. 
Then, the equations (\ref{3-10}) read 
%%%%%%%%%%%%%%%%%%%%%%%%%%%%%%%%%%%%%%%%%%%%%%%%%%%%%%%%%%%%%%%%
\begin{eqnarray}\label{3-10a}
  & &\left(-m^2 + 4\lambda {\overline \varphi}_{0}^2
     +12\lambda \int\!\frac{d^3{\mib k}}{(2\pi)^3}
       \ {\eta_{\bf k}^{0}}^2
%       \nonumber\\
%  & & \qquad\qquad\qquad\qquad
     +12\lambda \int\!\frac{d^3{\mib k}}{(2\pi)^3}\ 
     {\eta_{\bf k}^{\pi}}^2 \right)
     {\overline \varphi}_{0}-h=0 \ , \nonumber\\
  & &\left( {\mib k}^2+M_{\sigma}^2\right) \eta_{\bf k}^{0}
  =\frac{1}{4{\eta_{\bf k}^{0}}^{3}} \ , \nonumber\\
  & &\left( {\mib k}^2+M_{\pi}^2\right) \eta_{\bf k}^{\pi}
  =\frac{1}{4{\eta_{\bf k}^{\pi}}^{3}} \ ,
\end{eqnarray}
%%%%%%%%%%%%%%%%%%%%%%%%%%%%%%%%%%%%%%%%%%%%%%%%%%%%%%%%%%%%%%%
where we define the sigma meson mass $M_{\sigma}$ and 
the pion mass $M_{\pi}$ as 
%%%%%%%%%%%%%%%%%%%%%%%%%%%%%%%%%%%%%%%%%%%%%%%%%%%%%%%%%%%%%%%
\begin{eqnarray}\label{3-10b}
  & & M_{\sigma}^2
  =-m^2+12\lambda {\overline \varphi}_{0}^2 
    + 12\lambda\int\!\frac{d^3{\mib k}'}{{(2\pi)^3}}\  
    {\eta_{{\bf k}'}^{0}}^2 
%    \nonumber\\
%  & & \qquad\qquad\qquad\qquad
     +12\lambda \int\!\frac{d^3{\mib k}'}{(2\pi)^3}\ 
     {\eta_{{\bf k}'}^{\pi}}^2 \ , \nonumber\\
& &M_{\pi}^2
    =-m^2+4\lambda {\overline \varphi}_{0}^2 
    + 4\lambda\int\!\frac{d^3{\mib k}'}{{(2\pi)^3}}\  
    {\eta_{{\bf k}'}^{0}}^2 
%    \nonumber\\
%  & & \qquad\qquad\qquad\qquad
     +20\lambda \int\!\frac{d^3{\mib k}'}{(2\pi)^3}\ 
     {\eta_{{\bf k}'}^{\pi}}^2 \ .
\end{eqnarray}
%%%%%%%%%%%%%%%%%%%%%%%%%%%%%%%%%%%%%%%%%%%%%%%%%%%%%%%%%%%%
This is identical with the results derived in the previous 
paper.\cite{TVM1} (The parameters $\lambda/24$ and $m_0^2$ 
should be replaced into $\lambda$ and $-m^2$, respectively, 
in this paper. ) 
From Eq.(\ref{3-10a}), ${\eta_{\bf k}^0}^2$ and 
${\eta_{\bf k}^{\pi}}^2$ are expressed as 
%%%%%%%%%%%%%%%%%%%%%%%%%%%%%%%%%%%%%%%%%%%%%%%%%%%%%%%%%%%%%
\begin{equation}\label{3-10c}
{\eta_{\bf k}^0}^2=\frac{1}{2\sqrt{{\mib k}^2+M_\sigma^2}} \ , 
\qquad
{\eta_{\bf k}^{\pi}}^2=\frac{1}{2\sqrt{{\mib k}^2+M_\pi^2}} \ 
\end{equation}
and the two point function $G_0$ and $G_{\pi}$ are also obtained as
\begin{equation}\label{3-10d}
G_0=\int\!\frac{d^3{\mib k}}{(2\pi)^3}
\frac{1}{2\sqrt{{\mib k}^2+M_\sigma^2}} \ , 
\qquad
G_{\pi}=\int\!\frac{d^3{\mib k}}{(2\pi)^3}
\frac{1}{2\sqrt{{\mib k}^2+M_\pi^2}} \ .
\end{equation}
%%%%%%%%%%%%%%%%%%%%%%%%%%%%%%%%%%%%%%%%%%%%%%%%%%%%%%%%%%%%%
The equations (\ref{3-10a})$\sim$(\ref{3-10c}) compose 
the set of self-consistent equations.

\section{Numerical result and discussion}

We investigate the time-evolution of the mean field and 
quantum fluctuations around it based on the above-derived equations 
of motion in the context of the quench scenario in the relativistic 
nucleus-nucleus collision as was mentioned in \S 1. 
Our attention is paid to describe the damping of 
the amplitude of the mean field. It will be shown numerically 
that the fluctuation modes are responsible for the damping of 
the mean field configuration. Namely, since the energy stored in the 
mean field configuration flows to the fluctuation modes, 
the behavior like the damped oscillation appears in the mean field 
configuration. 
Our aim is here to describe the damping behavior qualitatively.

In general, 
the renormalization procedure is necessary in our approach 
because the loop contribution is included through the function 
$G({\mib x},{\mib x},t)$.
The renormalization will be carried out because the linear sigma 
model is renormalizable. 
However, 
it should be noted that 
the sigma model is here regarded as a low energy effective 
model of QCD. 
Thus, we become free from the complexity of the renormalization 
due to introducing the momentum cutoff in the context of the low 
energy effective theory.

%%%%%%%%%%%%%%%%%%%%%%%%%%%%% (Fig.1) %%%%%%%%%%%%%%%%%%%%%%%%%%%%%%%
\begin{figure}[t]
  \epsfxsize=7cm  % \epsfysize=   cm
  \centerline{\epsfbox{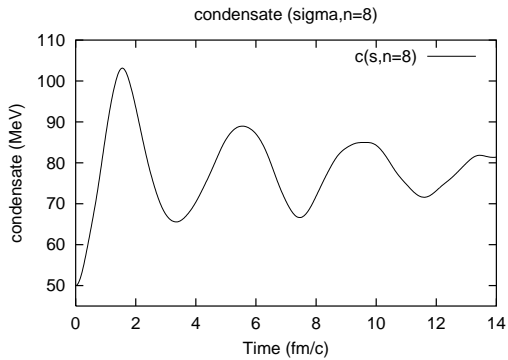}}
  \caption{The time-evolution of the chiral condensate is depicted 
  in which the fluctuation modes are contained up to $n^2=8^2$.}
   \label{fig:1}
\end{figure}
%%%%%%%%%%%%%%%%%%%%%%%%%%%%%%%%%%%%%%%%%%%%%%%%%%%%%%%%%%%%%%%%%%%%
%%%%%%%%%%%%%%%%%%%%%%%%%%%%% (Fig.2) %%%%%%%%%%%%%%%%%%%%%%%%%%%%%%%
\begin{figure}[t]
  \epsfxsize=6.5cm  % \epsfysize=   cm
  \epsfbox{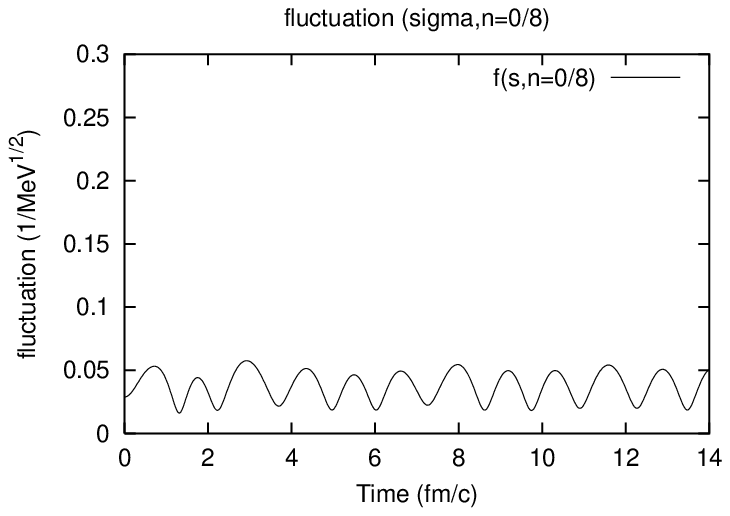}
   \hfill
  \epsfxsize=6.5cm  % \epsfysize=   cm
  \epsfbox{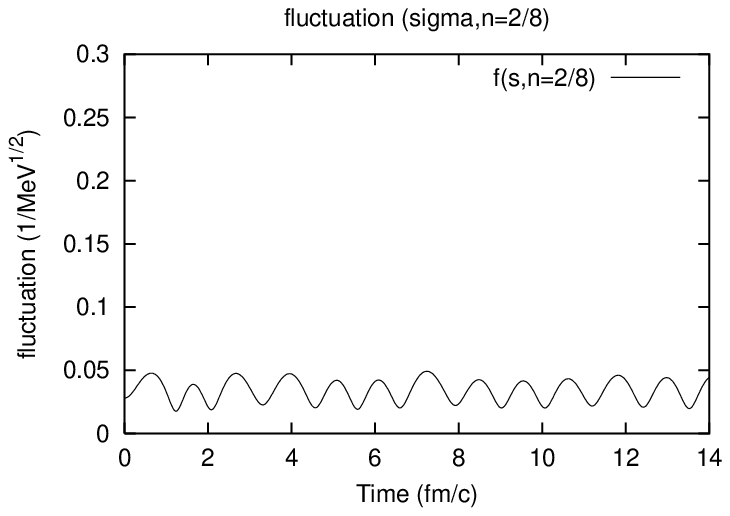}
  \epsfxsize=6.5cm  % \epsfysize=   cm
  \epsfbox{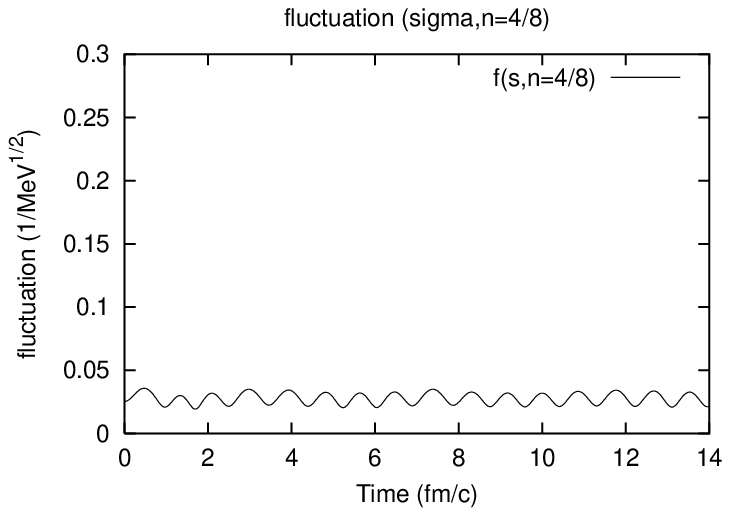}
   \hfill
  \epsfxsize=6.5cm  % \epsfysize=   cm
  \epsfbox{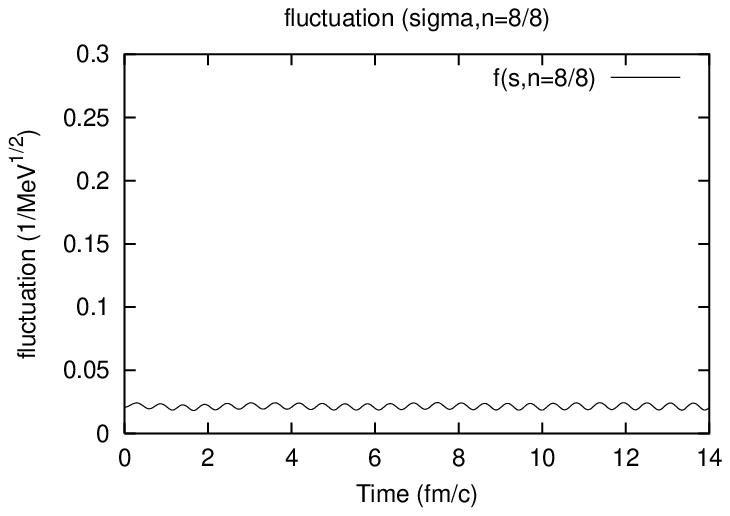}
  \caption{The time-evolution of the fluctuation modes in the 
  $\sigma$-direction are depicted. 
  The figures of the upper-left, upper-right, lower-left and lower-right 
  represent the time-evolution of the fluctuation modes with the quantum 
  numbers $n^2=0$, $n^2=2^2$, $n^2=4^2$ and $n^2=8^2$, respectively.}
   \label{fig:2}
\end{figure}
%%%%%%%%%%%%%%%%%%%%%%%%%%%%%%%%%%%%%%%%%%%%%%%%%%%%%%%%%%%%%%%%%%%%

The model parameters are given so as to 
reproduce the pion mass ($M_\pi=138$ MeV), the sigma meson mass 
($M_\sigma = 600$ MeV) and the pion decay constant ($f_{\pi}=92$ 
MeV) in the static case, respectively, that is, 
%%%%%%%%%%%%%%%%%%%%%%%%%%%%%%%%%%%%%%%%%%%%%%%%%%%%%%%%%%%%%%%%
\begin{eqnarray}\label{4-0}
& &m^2=\frac{M_\sigma^2-3M_\pi^2}{2}
+\frac{3(M_\sigma^2-M_\pi^2)G_\pi}{f_\pi^2+G_0-G_\pi} \ , \nonumber\\
& &\lambda=\frac{M_\sigma^2-M_\pi^2}{8(f_\pi^2+G_0-G_\pi)} \ , 
\nonumber\\
& &h=\frac{M_\pi^2 f_\pi^3+M_\sigma^2 f_{\pi}(G_0-G_\pi)}
{f_\pi^2+G_0-G_\pi} \ ,
\end{eqnarray}
%%%%%%%%%%%%%%%%%%%%%%%%%%%%%%%%%%%%%%%%%%%%%%%%%%%%%%%%%%%%%%%%%
where $G_0$ and $G_\pi$ have been defined in Eq.(\ref{3-10d}) 
and can be calculated with an appropriate three-momentum cutoff. 
%$m^2=(m_\sigma^2-3m_\pi^2)/2=(352 {\rm MeV})^2$, 
%$\lambda = (m_\sigma^2-m_\pi^2)/(8f_\pi^2)=4.15$ and 
%$h=f_\pi m_\pi^2 = (119 {\rm MeV})^3$ in the O(4) sigma model. 
%By following the method developed in the previous work\cite{TFKY} 
%in order to choose the initial conditions, 
In general, the initial condition of the mean field is taken 
arbitrary in our framework. 
However, the quantum fluctuation modes 
should be taken so that the quantum effects are 
as small as possible in static case. 
Thus, we adopt the initial conditions of the quantum fluctuation modes 
so as to satisfy $\partial E_{\rm fl}/\partial\eta_{\bf k}^a=0$ 
with ${\overline \varphi}_0=92$ MeV, where $E_{\rm fl}$ represents 
the energy of the quantum fluctuation part. 
Also, we take ${\dot \eta}_{\bf k}^{a}(t=t_0)=0$.

Let us investigate the time-evolution of the mean 
field on the $\sigma$-direction only. 
We assume ${\overline \varphi}_{i}=0$ with $i=1 \sim 3$.
In numerical calculation, we adopt the box normalization with 
the spatial length being $L$ for each direction. 
We then impose the periodic boundary condition 
for the fluctuation modes, namely, the allowed values of momenta are 
$k_x=(2\pi/L)n_x$ and so on, where $n_x$ is integer. 
The fluctuation modes labeled by $(n_x, n_y, n_z)$ are 
included in each isospin direction up to $n^2=n_x^2+n_y^2+n_z^2=8^2$. 
This corresponds to the momentum cutoff $\Lambda\sim 1$ GeV (990 MeV) 
since we have adopted the collisional region as $L^3=10^3$ fm$^3$. 
It should be here noted that 
the equations of motion in Eq.(\ref{3-10}) has 
the time-reversal invariance. Thus, the amplitude of the oscillation 
of the chiral condensate becomes large again even if the behavior of 
the damped oscillation is realized. 
However, in the 
realistic situation, the quantized pions and/or sigma mesons 
are emitted with a certain energy. 
Therefore, in our approach, it is enough to investigate 
the behavior of the condensate before the amplitude increases 
again.

%%%%%%%%%%%%%%%%%%%%%%%%%%%%% (Fig.3) %%%%%%%%%%%%%%%%%%%%%%%%%%%%%%%
\begin{figure}[t]
  \epsfxsize=6.5cm  % \epsfysize=   cm
  \epsfbox{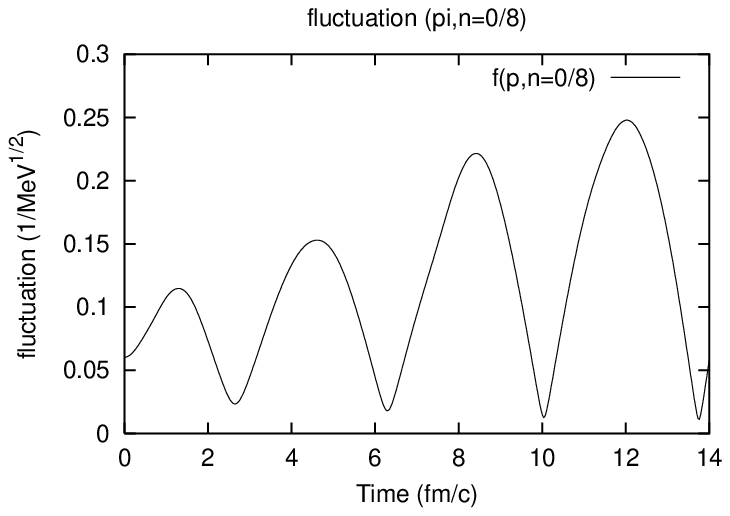}
   \hfill
  \epsfxsize=6.5cm  % \epsfysize=   cm
  \epsfbox{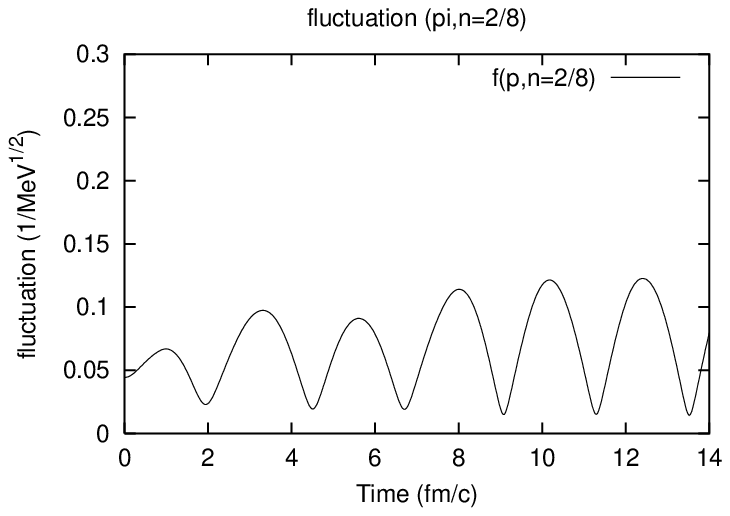}
  \epsfxsize=6.5cm  % \epsfysize=   cm
  \epsfbox{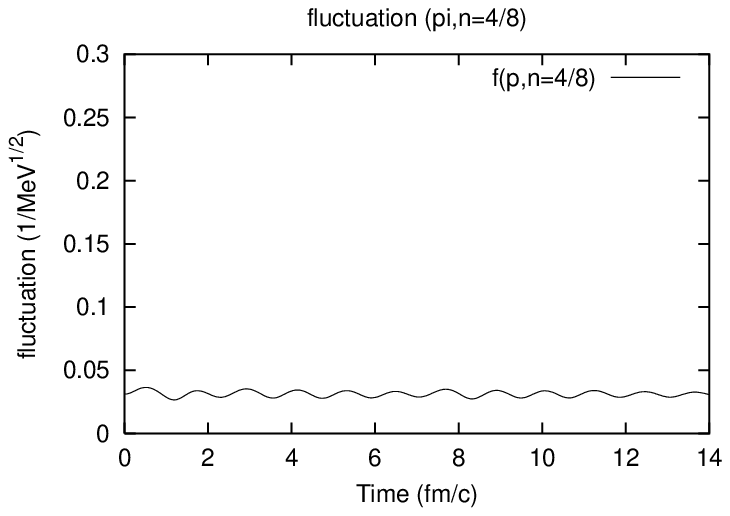}
   \hfill
  \epsfxsize=6.5cm  % \epsfysize=   cm
  \epsfbox{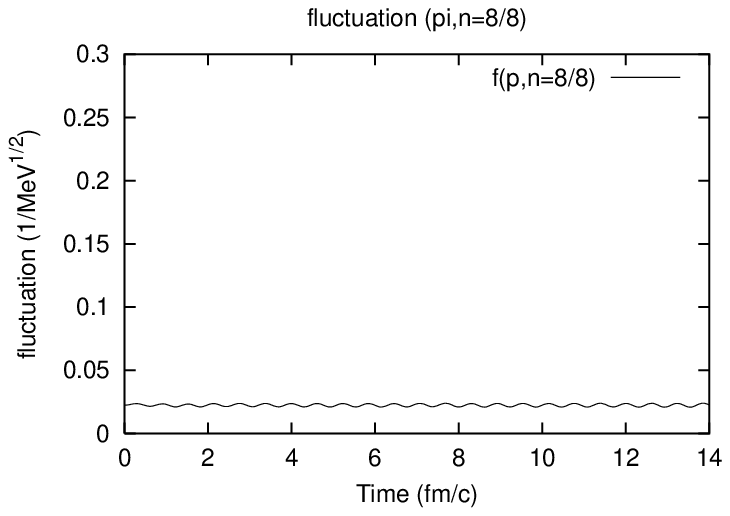}
  \caption{The time-evolution of the fluctuation modes in 
  the $\pi$-direction are depicted. 
  The figures of the upper-left, upper-right, lower-left and lower-right 
  represent the time-evolution of the fluctuation modes with the quantum 
  numbers $n^2=0$, $n^2=2^2$, $n^2=4^2$ and $n^2=8^2$, respectively.}
   \label{fig:3}
\end{figure}
%%%%%%%%%%%%%%%%%%%%%%%%%%%%%%%%%%%%%%%%%%%%%%%%%%%%%%%%%%%%%%%%%%%%

%%%%%%%%%%%%%%%%%%%%%%%%%%%%% (Fig.4) %%%%%%%%%%%%%%%%%%%%%%%%%%%%%%%
\begin{figure}[t]
  \epsfxsize=7cm  % \epsfysize=   cm  
  \centerline{\epsfbox{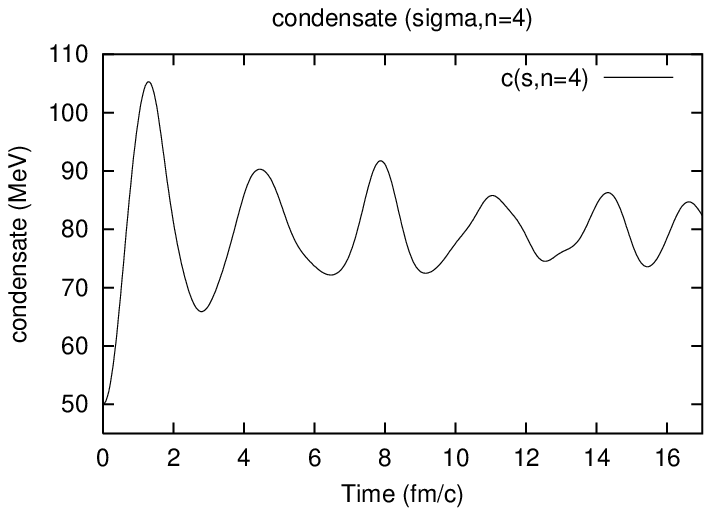}}
  \caption{The time-evolution of the chiral condensate is depicted 
  in which the fluctuation modes are contained up to $n^2=4^2$.}
   \label{fig:4}
\end{figure}
%%%%%%%%%%%%%%%%%%%%%%%%%%%%%%%%%%%%%%%%%%%%%%%%%%%%%%%%%%%%%%%%%%%%
%%%%%%%%%%%%%%%%%%%%%%%%%%%%% (Fig.5) %%%%%%%%%%%%%%%%%%%%%%%%%%%%%%%
\begin{figure}[t]
  \epsfxsize=6.5cm  % \epsfysize=   cm
  \epsfbox{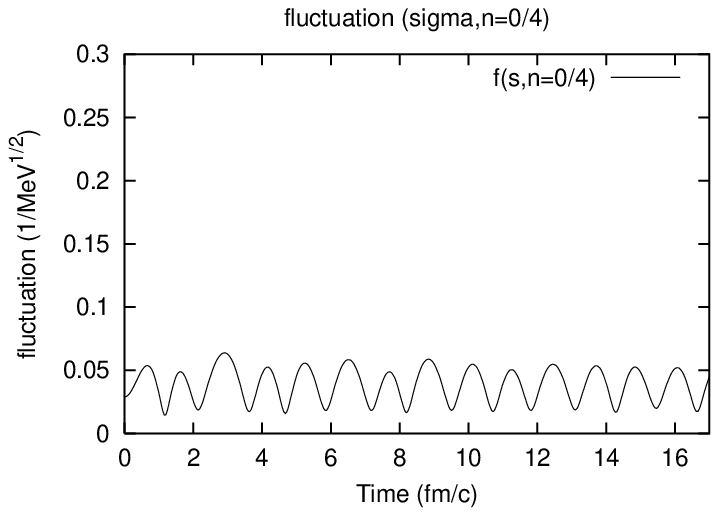}
   \hfill
  \epsfxsize=6.5cm  % \epsfysize=   cm
  \epsfbox{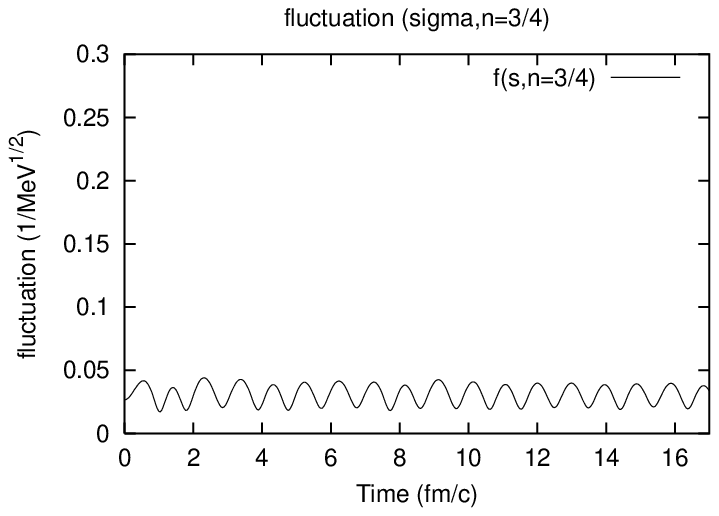}
  \caption{
  %The time-evolution of the fluctuation modes in the 
  %$\sigma$-direction are depicted. 
  %The left and the right figures 
  %represent 
  The time-evolution of the fluctuation modes with the quantum 
  numbers $n^2=0$ (left) and $n^2=3^2$ (right) 
  are depicted in the $\sigma$-direction.}
   \label{fig:5}
\end{figure}
%%%%%%%%%%%%%%%%%%%%%%%%%%%%%%%%%%%%%%%%%%%%%%%%%%%%%%%%%%%%%%%%%%%%
%%%%%%%%%%%%%%%%%%%%%%%%%%%%% (Fig.6) %%%%%%%%%%%%%%%%%%%%%%%%%%%%%%%
\begin{figure}[t]
  \epsfxsize=6.5cm  % \epsfysize=   cm
  \epsfbox{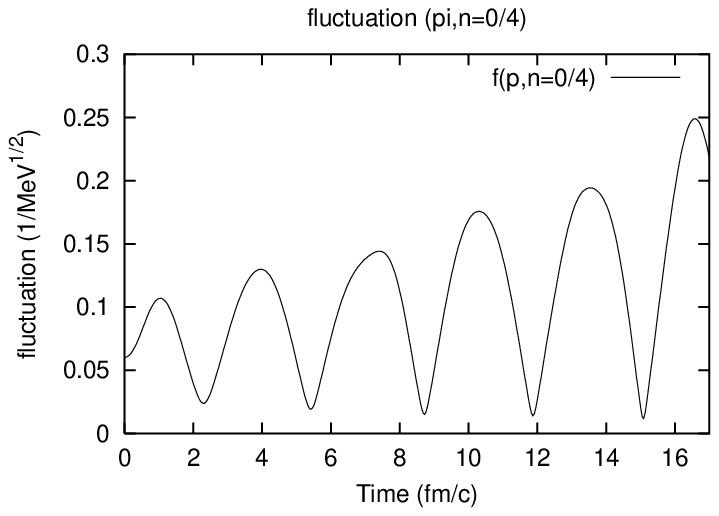}
   \hfill
  \epsfxsize=6.5cm  % \epsfysize=   cm
  \epsfbox{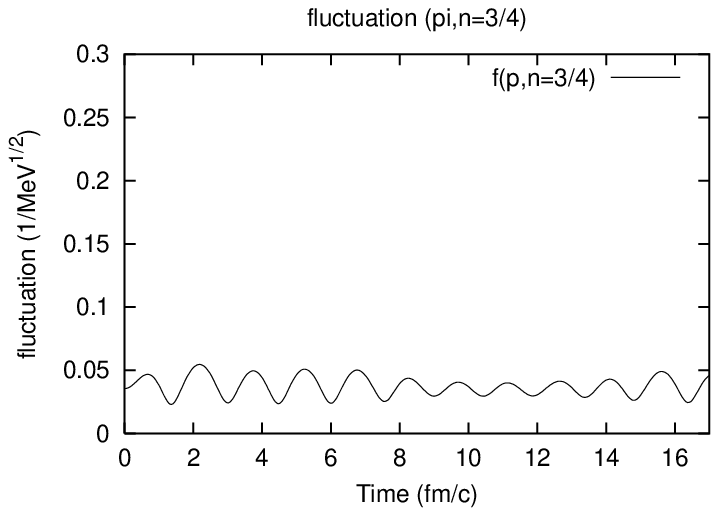}
  \caption{
  %The time-evolution of the fluctuation modes in the 
  %$\pi$-direction are depicted. 
  %The left and the right figures 
  %represent 
  The time-evolution of the fluctuation modes with the quantum 
  numbers $n^2=0$ (left) and $n^2=3^2$ (right) are depicted 
  in the $\pi$-direction.
  }
   \label{fig:6}
\end{figure}
%%%%%%%%%%%%%%%%%%%%%%%%%%%%%%%%%%%%%%%%%%%%%%%%%%%%%%%%%%%%%%%%%%%%

Figure 1 shows us the time-evolution of the chiral condensate 
with ${\overline \varphi}(t=t_0=0)=50$ MeV. 
It is shown that the behavior of the damped oscillation is 
realized. It should be noted that the center value of oscillation is 
slightly different from 92 MeV. The reason is that 
the effective potential ${\cal M}^{(0)}[{\overline \varphi}]$ 
contains the effects of quantum fluctuations in our approach. 
If the quantum fluctuation modes have no time-dependence, the 
value of condensate should be 92 MeV. 
It is seen in Fig.2 that the $\sigma$-modes oscillate. 
However, in Fig.3, it is seen that 
the amplitude of $\pi$-modes with low momenta ($n^2\le 2^2$) 
increases. 
Namely, these amplitude of 
$\pi$-modes with low momenta increases corresponding to 
the decreasing of 
the amplitude of the chiral condensate. 
However, the $\pi$-mode with high momenta are not amplified but 
oscillated only. 
This situation can be easily understood. 
The instability occurs when the time-dependent ``mass" term 
$M_a(t)^2=-m^2+12\lambda
{\overline \varphi}_a(t)+\sum_{b\neq a}4\lambda{\overline \varphi}_b(t)
+(\hbox{\rm quantum\ correction})$ 
is negative and the absolute value of it is greater than ${\mib k}^2$. 
For $\sigma$-mode, the time-dependent mass is 
$M_0(t)^2=-m^2+12\lambda{\overline \varphi}_0(t)
+(\hbox{\rm quantum\ correction})$ as is similar to Eq.(\ref{3-10b}). 
However, for the $\pi$-mode, the mass $M_\pi(t)^2$ is 
less than $M_0(t)^2$ 
because $M_\pi(t)^2=-m^2+4\lambda{\overline \varphi}_0(t)
+(\hbox{\rm quantum\ correction})$. Thus, the 
instability occurs in the $\pi$-modes with low momenta. 
As a result, the behavior of the 
amplified oscillation is realized. 
On the other word, as was shown in \S 2, 
if the mode functions are expressed as 
$\eta_{\bf k}(t)=1/\sqrt{2\epsilon_{\bf k}(t)}$, 
the instability is realized when $\epsilon_{\bf k}^2(t)<0$. 
When $\epsilon_{\bf k}^2(t)<0$ for $|{\mib k}|<\sqrt{-M^2}$, 
then $G({\mib x},{\mib x},t)$ 
is obtained as 
%%%%%%%%%%%%%%%%%%%%%%%%%%%%% (Fig.7) %%%%%%%%%%%%%%%%%%%%%%%%%%%%%%%
\begin{figure}[t]
  \epsfxsize=7cm  % \epsfysize=   cm
  \centerline{\epsfbox{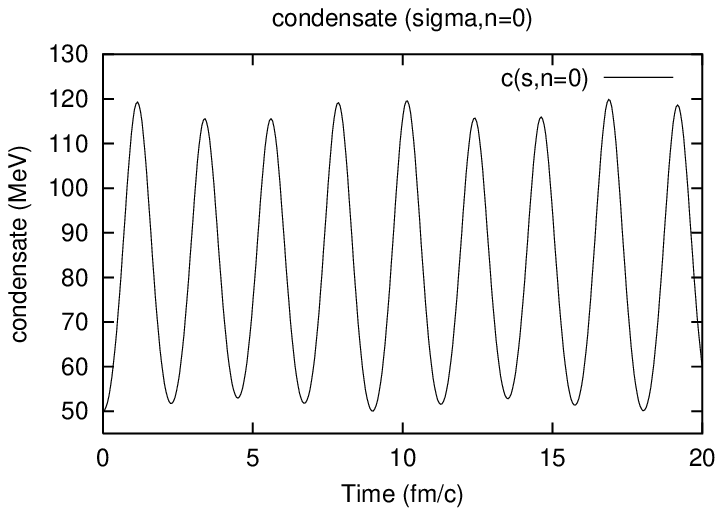}}
  \caption{The time-evolution of the chiral condensate is depicted 
  in which the fluctuation mode with $n=0$ is only contained.}
   \label{fig:7}
\end{figure}
%%%%%%%%%%%%%%%%%%%%%%%%%%%%%%%%%%%%%%%%%%%%%%%%%%%%%%%%%%%%%%%%%%%%
%%%%%%%%%%%%%%%%%%%%%%%%%%%%% (Fig.8-9) %%%%%%%%%%%%%%%%%%%%%%%%%%%%%%%
\begin{figure}[t]
 \parbox{\halftext}{%   %\def\halftext{.6\textwidth}
  \epsfxsize=6.5cm  % \epsfysize=   cm
  \epsfbox{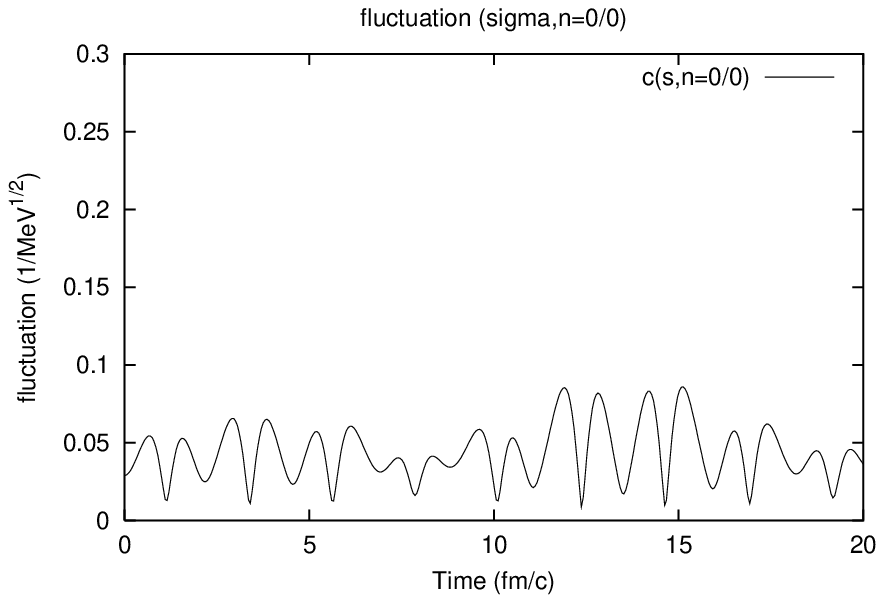}
  \caption{The time-evolution of the fluctuation mode with $n=0$ 
  in the $\sigma$-direction is depicted. 
  }
   \label{fig:8}}
  \hfill
 \parbox{\halftext}{
  \epsfxsize=6.5cm  % \epsfysize=   cm
  \epsfbox{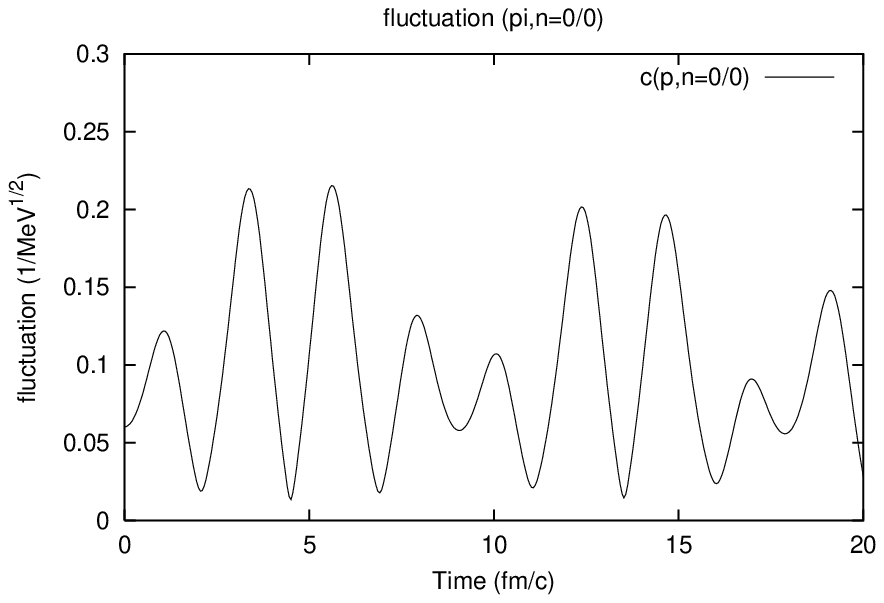}
  \caption{The time-evolution of the fluctuation mode with $n=0$ 
  in the $\pi$-direction is depicted. }
   \label{fig:9}}
\end{figure}
%%%%%%%%%%%%%%%%%%%%%%%%%%%%%%%%%%%%%%%%%%%%%%%%%%%%%%%%%%%%%%%%%%%%
%
%%%%%%%%%%%%%%%%%%%%%%%%%%%%%%%%%%%%%%%%%%%%%%%%%%%%%%%%%%%%%%%
\begin{equation}\label{4-1}
G({\mib x},{\mib x},t)=\sum_{|{\bf k}|<\sqrt{-M^2}}
\frac{-i}{2|\epsilon_{\bf k}(t)|}
e^{\pm2\int^t |\epsilon_{\bf k}(t')|dt'}
+\sum_{|{\bf k}|>\sqrt{-M^2}}\frac{1}{2\epsilon_{\bf k}(t)} \ .
\end{equation}
Namely, this means that 
the imaginary part appears in the equation of motion for 
the mean field ${\overline \varphi}_a(t)$. 
In this case the amplitudes of the low momentum modes with 
$|{\mib k}|<\sqrt{-M^2}$ 
of quantum fluctuation increase in time. 
Thus, the relaxation process for the chiral condensate occurs 
due to the instability of the quantum fluctuation. 
Then, the energy stored in the chiral condensate flows to 
the quantum fluctuations.

%%%%%%%%%%%%%%%%%%%%%%%%%%%%% (Fig.10) %%%%%%%%%%%%%%%%%%%%%%%%%%%%%%%
\begin{figure}[t]
  \epsfxsize=7cm  % \epsfysize=   cm
  \centerline{\epsfbox{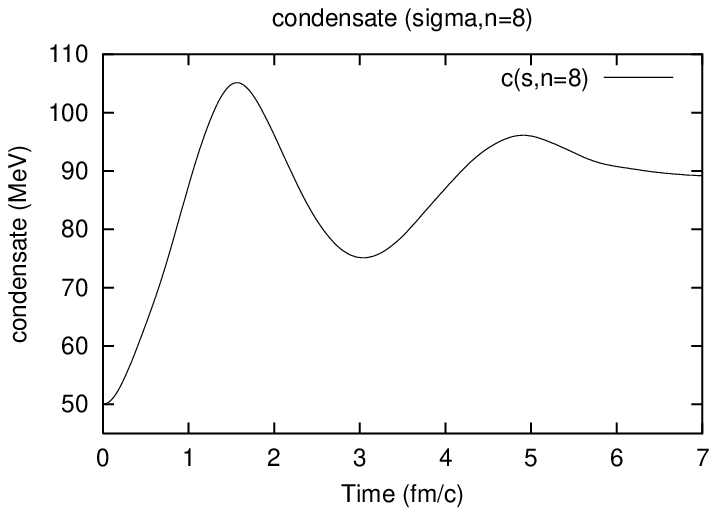}}
  \caption{The time-evolution of the chiral condensate on the 
  expanding geometry is depicted 
  in which the fluctuation modes are contained up to $n^2=8^2$.}
   \label{fig:10}
\end{figure}
%%%%%%%%%%%%%%%%%%%%%%%%%%%%%%%%%%%%%%%%%%%%%%%%%%%%%%%%%%%%%%%%%%%%
%%%%%%%%%%%%%%%%%%%%%%%%%%%%% (Fig.11) %%%%%%%%%%%%%%%%%%%%%%%%%%%%%%%
\begin{figure}[t]
  \epsfxsize=6.5cm  % \epsfysize=   cm
  \epsfbox{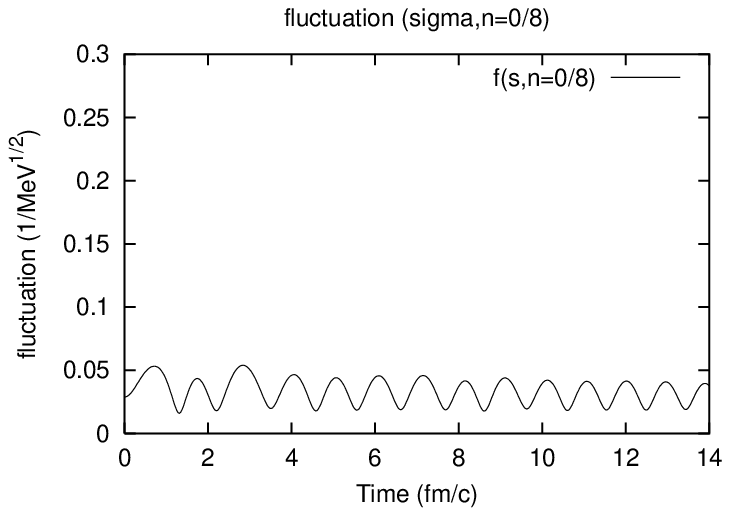}
   \hfill
  \epsfxsize=6.5cm  % \epsfysize=   cm
  \epsfbox{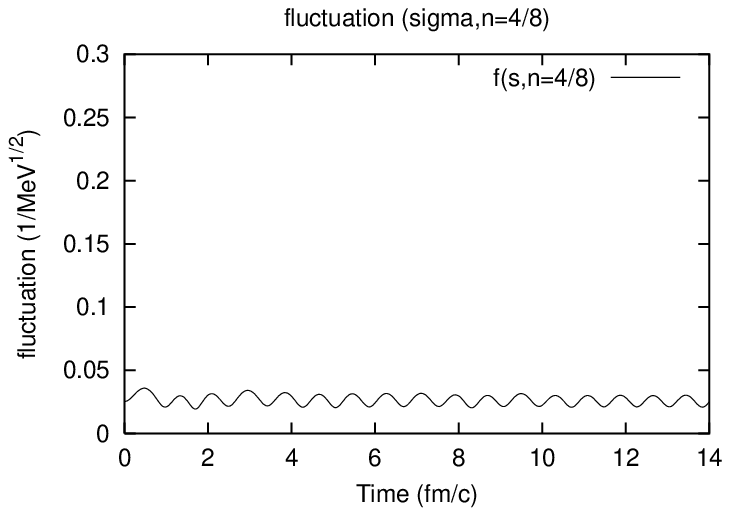}
  \caption{The time-evolution of the fluctuation modes 
  with the quantum number $n^2=0$ (left) and $n^2=4^2$ (right) in the 
  $\sigma$-direction on the expanding geometry are depicted. 
  %The left and the right figures 
  %represent the time-evolution of the fluctuation modes with the quantum 
  %numbers $n^2=0$ and $n^2=4^2$, respectively.
  }
   \label{fig:11}
\end{figure}
%%%%%%%%%%%%%%%%%%%%%%%%%%%%%%%%%%%%%%%%%%%%%%%%%%%%%%%%%%%%%%%%%%%%
%%%%%%%%%%%%%%%%%%%%%%%%%%%%% (Fig.12) %%%%%%%%%%%%%%%%%%%%%%%%%%%%%%%
\begin{figure}[t]
  \epsfxsize=6.5cm  % \epsfysize=   cm
  \epsfbox{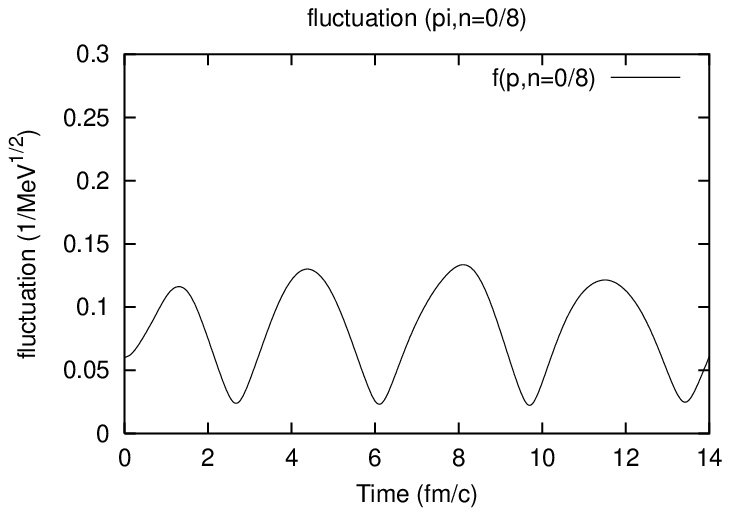}
   \hfill
  \epsfxsize=6.5cm  % \epsfysize=   cm
  \epsfbox{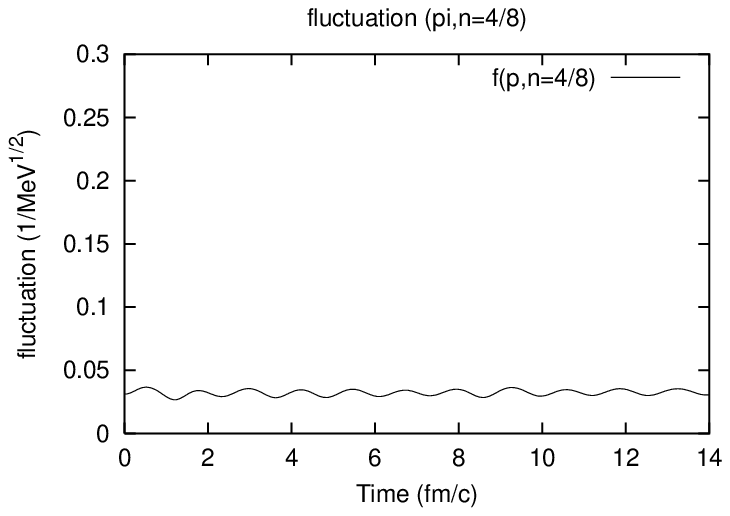}
  \caption{The time-evolution of the fluctuation modes 
  with the quantum number $n^2=0$ (left) and $n^2=4^2$ (right) 
  in the 
  $\pi$-direction on the expanding geometry are depicted. 
  %The left and the right figures 
  %represent the time-evolution of the fluctuation modes with the quantum 
  %numbers $n^2=0$ and $n^2=4^2$, respectively.
  }
   \label{fig:12}
\end{figure}
%%%%%%%%%%%%%%%%%%%%%%%%%%%%%%%%%%%%%%%%%%%%%%%%%%%%%%%%%%%%%%%%%%%%

In Figs.4$\sim$6 and 7$\sim$9, 
the only a few fluctuation modes up to 
$n^2=4^2$ and $n^2=0$ are 
taken into account in each isospin direction, respectively. 
The time-evolution of the condensate, the fluctuation modes 
with $\sigma$-direction and the fluctuation modes with 
$\pi$-directions 
are depicted in Figs.4, 5 and 6, respectively, up to $n^2=4^2$. 
The qualitative behavior is almost same as that in Figs.1$\sim$3. 
The fluctuation modes with $n^2\ge 3^2$ 
for $\pi$-mode are not responsible for the damping of the amplitude 
of the condensate. However, in Figs.7$\sim$9, 
the qualitative behavior 
is quite different from Figs.1$\sim$3 and 4$\sim$6. 
The condensate is relaxed 
to the vacuum value no longer. The reason why the relaxation process 
is not realized is that the quantum fluctuation modes are not 
included sufficiently. Thus we conclude that the quantum fluctuation 
mode for $\pi$-direction with low momenta are responsible for the 
relaxation process of the chiral condensate during the chiral 
phase transition.

Let us consider the expanding geometry\cite{DOMINIQUE} 
to break the time-reversal invariance. 
The basic equations are given in Appendix. 
The expanding parameter $a(t)$ is assumed so as to be 
proportional to time : 
$a(t)=t/\tau_0+1$. 
The parameter is taken as $\tau_0=1$ fm/$c$. The other 
parameters are same as those in Figs.1$\sim$3. 
Also, it is assumed that 
the expansion occurs in $D=3$-dimension. 
Figure 10 shows us that the behavior of the damped oscillation 
appears in the chiral condensate and the oscillation converges 
to the vacuum value. It is also seen that 
the amplitude of the 
quantum fluctuation modes also decreases gradually in time 
in Fig.11 ($\sigma$-modes) and Fig.12 ($\pi$-modes).

\section{Summary}

In this paper, we have investigated the 
time-evolution of the chiral condensate and the quantum fluctuation 
around it in the context of the chiral phase transition from the 
symmetric to symmetry-broken phase. 
We have formulated the time-dependent variational method 
with the squeezed state for O(N) linear sigma model. 
It has been shown numerically that 
the behavior like the damped oscillation for 
the chiral condensate appears due to the inclusion of the 
quantum fluctuations in the squeezed state approach. 
We have pointed out that the quantum 
fluctuations, especially $\pi$-modes with low momenta, are 
responsible for the occurrence of the relaxation process of 
the chiral condensate. 
Of course, the expansion of the system 
leads to a friction term.\cite{RUNDRAP,ISHIHARA} 
However, even if the expansion is not taken 
into account, it is concluded that 
the relaxation process occurs due to 
the quantum fluctuations around the mean field configuration.
It may be also interesting to investigate the implication to 
the parametric resonance in the $\pi$-mode.

\section*{Acknowledgements}

One of the authors (Y.T.) would like to express his sincere 
thanks to Professor\break
Dominique Vautherin for giving his interest 
in this work and stimulating discussions when he stayed in 
the Laboratoire de Physique Th\'eorique des Particules 
El\'ementaires (LPTPE) of Universit\'e Pierre et Marie
Curie (Paris VI). 
He also wishes to thank Dr. H. Fujii in University of Tokyo 
for the early collaboration in this work 
and giving him useful comments.

\appendix
\section{Expanding geometry} 
%Empty argument \section{} yields `Appendix'. 
%
The collisional region may expand spatially. 
If it is necessary to include the effect of the expanding 
geometry,\cite{DOMINIQUE} 
the metric may be taken as 
%
%%%%%%%%%%%%%%%%%%%%%%%%%%%%% (3.11) %%%%%%%%%%%%%%%%%%%%%%%%%%%%%%%
\begin{equation}\label{3-11}
   ds^2=dt^2-a(t)^2 d{\mib x}^2 \ \ =g_{\mu\nu}(t)dx^{\mu} dx^{\nu} \ ,
\end{equation}
%%%%%%%%%%%%%%%%%%%%%%%%%%%%%%%%%%%%%%%%%%%%%%%%%%%%%%%%%%%%%%%%%%
%
where $a(t)$ represents the expansion of collisional region, 
which is governed by another dynamics. 
Here, $a(t)$ is given by hand. 
Hereafter, it is assumed that 
the expansion occurs in $D=1$- or $D=3$-dimension. 
Further, in $D=1$, the direction of 
the expansion is adopted as the $z$-direction.
The metric tensor $g_{\mu\nu}(t)$, its inverse and its determinant $g$ 
are given by
%
%%%%%%%%%%%%%%%%%%%%%%%%%%%%% (A.1) %%%%%%%%%%%%%%%%%%%%%%%%%%%%%%%
\begin{eqnarray}\label{A-1}
   g_{\mu\nu}(t)
    &=& \pmatrix{
          1 & 0 & 0 & \cdots & 0 \cr
          0 & -a(t)^2 & 0 & \cdots & 0 \cr
          0 & 0 & \ddots & \cdots & 0 \cr
          \vdots & \vdots & \vdots & \ddots & 0 \cr
          0 & 0 & 0 & \cdots & -a(t)^2 \cr
                } \ , \nonumber\\
 & &\nonumber\\
    g^{\mu\nu}(t)
    &=& \pmatrix{
          1 & 0 & 0 & \cdots & 0 \cr
          0 & -1/a(t)^2 & 0 & \cdots & 0 \cr
          0 & 0 & \ddots & \cdots & 0 \cr
          \vdots & \vdots & \vdots & \ddots & 0 \cr
          0 & 0 & 0 & \cdots & -1/a(t)^2 \cr
                } \ , \nonumber\\
 & & \nonumber\\
   g &=& {\rm det}g_{\mu\nu} = (-a(t)^2)^D 
\end{eqnarray}
%%%%%%%%%%%%%%%%%%%%%%%%%%%%%%%%%%%%%%%%%%%%%%%%%%%%%%%%%%%%%%%%%%
%
with $D=1$ or 3. 
Then, equations of motion in Eq.(\ref{3-10}) are modified because 
the d'Alembertian 
$\partial^2/\partial t^2 - \nabla^2$, which acts on 
${\overline \varphi}_a$ and $\eta^a$, 
is replaced by 
%
%%%%%%%%%%%%%%%%%%%%%%%%%%%%% (A.2) %%%%%%%%%%%%%%%%%%%%%%%%%%%%%%%
\begin{eqnarray}\label{A-2}
   {\overline \varphi}_{a{;\mu}}^{\ ;\mu}
    &=& \frac{1}{\sqrt{-g}}\frac{\partial}{\partial x^{\mu}}
        \left(\sqrt{-g}g^{\mu\nu}
        \frac{\partial {\overline \varphi}_a}{\partial x^{\nu}}
        \right) \nonumber\\
    &=&\frac{1}{\sqrt{-g}}\left[ \frac{\partial}{\partial t}
       \left(\sqrt{-g}\frac{\partial{\overline \varphi}_a}
       {\partial t}\right)
       -\nabla\cdot\left(\sqrt{-g}\frac{1}{a(t)^2}\nabla
       {\overline \varphi}_a\right)
       \right] \nonumber\\
    &=&\frac{\partial^2 {\overline \varphi}_a}{\partial t^2}
       +D\frac{{\dot a}(t)}{a(t)}\frac{\partial {\overline \varphi}_a}
       {\partial t}
       -\nabla_D^2 {\overline \varphi}_a \ .
\end{eqnarray}
%%%%%%%%%%%%%%%%%%%%%%%%%%%%%%%%%%%%%%%%%%%%%%%%%%%%%%%%%%%%%%%%%%
%
Namely, the d'Alembertian is replaced into 
%
%%%%%%%%%%%%%%%%%%%%%%%%%%%%% (3.12) %%%%%%%%%%%%%%%%%%%%%%%%%%%%%%%
\begin{equation}\label{3-12}
   \frac{\partial^2}{\partial t^2}-\nabla^2 
   \longrightarrow 
   \frac{\partial^2}{\partial t^2}+D\frac{{\dot a}(t)}{a(t)}
   \frac{\partial}{\partial t}-\nabla_D \ , 
\end{equation}
%%%%%%%%%%%%%%%%%%%%%%%%%%%%%%%%%%%%%%%%%%%%%%%%%%%%%%%%%%%%%%%%%%
%
where the variable with suffix $D\ (=1, 3)$ is defined as
%
%%%%%%%%%%%%%%%%%%%%%%%%%%%%% (3.13) %%%%%%%%%%%%%%%%%%%%%%%%%%%%%%%
\begin{equation}\label{3-13}
   f_D^2=\frac{1}{a(t)^{D-1}}(f_x^2+f_y^2)+\frac{1}{a(t)^2}f_z^2 \ .
\end{equation}
%%%%%%%%%%%%%%%%%%%%%%%%%%%%%%%%%%%%%%%%%%%%%%%%%%%%%%%%%%%%%%%%%%
%
As a result, the equations of motion we solve are modified 
from Eq.(\ref{3-10}) as follows :
%
%%%%%%%%%%%%%%%%%%%%%%%%%%%%% (3-14) %%%%%%%%%%%%%%%%%%%%%%%%%%%%%%%
\begin{eqnarray}\label{3-14}
  & &{\ddot {\overline \varphi}}_{a}(t)+D\frac{{\dot a}(t)}{a(t)}
  {\dot {\overline \varphi}}_{a}(t)
     -m^2 {\overline \varphi}_{a}(t) 
     + 4\lambda {\overline \varphi}_{a}(t)^3
     +12\lambda \int\!\frac{d^3{\mib k}}{(2\pi)^3}
       \ \eta_{\bf k}^{a}(t)^2\cdot 
       {\overline \varphi}_{a}(t) \nonumber\\
  & & \qquad\qquad\qquad\qquad
     +4\lambda \sum_{b\neq a}\left({\overline \varphi}_{b}(t)^2 
     +\int\!\frac{d^3{\mib k}}{(2\pi)^3}\ \eta_{\bf k}^{b}(t)^2 \right)
     {\overline \varphi}_{a}(t) -h\delta_{a0} =0 \ , \nonumber\\
  & & {\ddot \eta}_{\bf k}^{a}(t) +D\frac{{\dot a}(t)}{a(t)}
  {\dot \eta}_{\bf k}^{a}(t) 
    + \biggl[{k}_D^2-m^2+12\lambda {\overline \varphi}_{a}(t)^2 
    + 12\lambda\int\!\frac{d^3{\mib k}'}{{(2\pi)^3}}\  
    \eta_{{\bf k}'}^{a}(t)^2 
       \nonumber\\
  & & \qquad\qquad\qquad\qquad
     +4\lambda \sum_{b\neq a}\left({\overline \varphi}_{b}(t)^2 
     +\int\!\frac{d^3{\mib k}'}{(2\pi)^3}\ \eta_{{\bf k}'}^{b}(t)^2 
     \right) 
       \biggl]
    \eta_{\bf k}^{a}(t) 
    -\frac{1}{4 \eta_{\bf k}^{a}(t)^3} = 0 \ . \nonumber\\
\end{eqnarray}
%%%%%%%%%%%%%%%%%%%%%%%%%%%%%%%%%%%%%%%%%%%%%%%%%%%%%%%%%%%%%%%%%%
%
Of course, the term with the first derivative of time 
causes the friction and the dissipative process is realized.

\end{document}